\documentclass[12pt]{article}
\usepackage{graphicx}
\usepackage{amsmath}

\begin{document}

\author{M. O. Hase (1) \and J. R. L. de Almeida (1,2) \and and S. R. Salinas (1) \\
(1) Instituto de F\'{\i}sica,\\
Universidade de S\~{a}o Paulo,\\
Caixa Postal 66318\\
05315-970, S\~{a}o Paulo, SP, Brazil\\
(2) Departamento de F\'{\i}sica\\
Universidade Federal de Pernambuco, \\
Recife, PE, Brazil}
\date{22 May 2005}
\title{Replica-symmetric solutions of a dilute Ising ferromagnet in a random field}
\maketitle

\begin{abstract}
We use the replica method in order to obtain an expression for the
variational free energy of an Ising ferromagnet on a Viana-Bray lattice in
the presence of random external fields. Introducing a global order
parameter, in the replica-symmetric context, the problem is reduced to the
analysis of the solutions of a nonlinear integral equation. At zero
temperature, and under some restrictions on the form of the random fields,
we are able to perform a detailed analysis of stability of the
replica-symmetric solutions. In contrast to the behaviour of the 
Sherrington-Kirkpatrick model for a spin glass in a uniform
field, the paramagnetic solution is fully stable in a sufficiently large
random field.


emails: mhase@if.usp.br, almeida@df.ufpe.br, ssalinas@if.usp.br.
\end{abstract}

\section{Introduction}

Magnetic systems with quenched disorder, including spin glasses and
ferromagnets in a random field, have been intensively studied during the
last decades \cite{Young1998}. There are many applications of these
disordered systems, ranging from the study of the behavior of random
magnets, which is a traditional ground test for the ideas of statistical
mechanics, to the analysis of different sorts of optimization problems in
distinct areas of science. The mean-field version of an Ising spin glass,
with Gaussian distribution of exchange interactions, also known as the
Sherrington-Kirkpatrick model, which can be solved by the replica method,
displays a low-temperature glassy phase, characterized by the instability of
a replica-symmetric solution, which indicates the need of breaking replica
symmetry and the existence of many ultrametrically organized states. In
contrast to this rich behavior, the mean-field, Curie-Weiss, version of an
Ising ferromagnet in a random field (RFIM), which can be solved without
recourse to the replica method, leads to rather uninteresting,
replica-symmetric, exact solutions. There have been, however, some
indications that a ferromagnet in a random field may have a glassy behavior.

We were motivated to look again at this problem by a number of early and
some more recent investigations of the RFIM. De Almeida and Bruinsma \cite
{Dealmeida1987} have done some calculations, beyond the usual Curie-Weiss,
mean-field, approximation, for analyzing the behavior of a bond-diluted
Ising antiferromagnet in a field. For large dimensionality, these
calculations lead to the presence of a glassy region in the applied field
versus temperature phase diagram, between paramagnetic and ordered phases,
which can be shown to disappear in the limit of infinite dimension. On the
basis of the equivalence, at a mean-field level, between the critical
behavior of a ferromagnet in a random field and of a dilute antiferromagnet
in a uniform field, this result gives an indication of the possible
existence of a glassy phase in the RFIM. A glassy behavior is also present
in a recent ``extended mean-field'' calculation by Pastor and collaborators 
\cite{Pastor2002} for the phase diagram of the RFIM. These results are
claimed to agree with earlier work of M\'{e}zard and Young \cite{Mezard1992}
using a screening approximation in order to characterize the instability of
the replica-symmetric solutions, to lowest order in $1/m$, in a
renormalization-group calculation for an $m$-component spin ferromagnet in a
random field. It should mentioned that calculations for the RFIM on a Bethe
lattice already indicate a rich ground-state structure \cite{Bruinsma1984}
and peculiar hysteresis effects \cite{Detcheverry}. Also, field-theoretical
renormalization-group calculations for a soft-spin version of the RFIM,
which were confirmed by a formulation of the dynamics, have shown the need
to include extra terms involving replicas with different indices, which in
turn may lead to an instability of the replica-symmetric solution in the
paramagnetic region \cite{Brezin1998}. The prediction of a glassy phase in
the calculations beyond the usual mean-field approximation were the main
motivation to revisit this problem. We then decided to use a model devised
by Viana and Bray \cite{Viana1985}, which is designed to gauge the effects
of the (finite) connectivity of a lattice.

According to the original work of Viana and Bray, we consider an Ising model
with pair interactions $J_{ij}$, between all sites $i$ and $j$, such that $%
J_{ij}=J>0$, with probability $c/N$, where $N$ is the total number of sites,
and $J_{ij}=0$, with probability $1-c/N$. This choice of interactions gives
rise to the so-called Viana-Bray lattice. The parameter $c>0$ can be
regarded as the (finite) mean connectivity per site. The solutions of an
Ising spin-glass on the Viana-Bray lattice have been first analyzed in the
vicinity of the transition temperature \cite{Viana1985}. The more involved
low-temperature behavior has been considered by Kanter and Sompolinsky \cite
{Kanter1987}, and M\'{e}zard and Parisi \cite{Mezard1987}. The analysis of
stability of the replica-symmetric solutions near percolation ($c=1$) has
been carried out by de Dominicis and collaborators \cite{Dedominicis1987}.
Although there are calculations for the RFIM on a Bethe lattice, there is no
comprehensive analysis on a graph as the Viana-Bray lattice. According to
the previous work for the Ising spin glass, we introduce a global order
parameter and formulate the replica-symmetric solutions for this problem in
terms of an integral equation. In the ground state, and under some
conditions on the form of the random fields, we show that the
replica-symmetric solutions can be written as a series expansion. We have
been able to perform a detailed analysis of stability of this
replica-symmetric solution. In a sufficiently strong random field, the
explicit calculation of the eigenvalues of a functional Hessian form shows
the stability of the paramagnetic solution, which seems to preclude the
existence of a glassy phase (in contrast to earlier expectations and the
results of Pastor and collaborators \cite{Pastor2002}). Some numerical
calculations confirm these findings for the paramagnetic solution, and
indicate that the replica-symmetric ferromagnetic solution is also stable in
the ferromagnetic region of the phase diagram.

The layout of this paper is as follows. In Section 2, we define the model
and formulate the replica-symmetric solutions in terms of an integral
equation for the global order parameter. The analysis of stability of the
paramagnetic solution is reported in Section 3. Some conclusions, and
connections with recent work, are presented in Section 4.

\section{Formulation of the problem}

The ferromagnetic Ising model on the Viana-Bray lattice is given by the
Hamiltonian 
\begin{equation}
H=-\sum_{\left( ij\right) }J_{ij}\sigma _{i}\sigma
_{j}-\sum_{i=1}^{N}H_{i}\sigma _{i},
\end{equation}
where $\left( ij\right) $ refers to a pair of sites, $\sigma _{i}=\pm 1$ for
all sites, and $\left\{ J_{ij}\right\} $ and $\left\{ H_{i}\right\} $ are
sets of independent, identically distributed, random variables, associated
with probability distributions 
\begin{equation}
p_{J}\left( J_{ij}\right) =\frac{c}{N}\delta \left( J_{ij}-J\right) +\left(
1-\frac{c}{N}\right) \delta \left( J_{ij}\right) ,
\end{equation}
and 
\begin{equation}
p_{H}\left( H_{i}\right) =\frac{1}{2}\delta \left( H_{i}-H_{R}\right) +\frac{%
1}{2}\delta \left( H_{i}+H_{R}\right) ,  \label{doubled}
\end{equation}
where $J$, $c$, and $H_{R}$ are positive parameters. This distribution of
exchange interactions is supposed to mimic a lattice of mean finite
connectivity $c$.

Using the replica method, it is not difficult to write the variational free
energy \cite{Dedominicis1987} 
\begin{equation*}
f=\frac{1}{\beta }\lim_{n\rightarrow 0}\frac{1}{n}\{\frac{c}{2}+\frac{c}{2}%
\sum_{r=0}^{n}\sum_{\left( \alpha _{1},...,\alpha _{r}\right)
}b_{r}q_{\alpha _{1},...,\alpha _{r}}^{2}-
\end{equation*}
\begin{equation}
-\ln \int_{-\infty }^{+\infty }dHp_{H}\left( H\right) \mbox{Tr}_{\sigma
}\exp \left[ G\left( \left\{ \sigma _{\alpha }\right\} \right) +\beta
H\sum_{\alpha =1}^{n}\sigma _{\alpha }\right] \},  \label{free}
\end{equation}
where $\beta =1/\left( k_{B}T\right) $, $T$ is the temperature, $n$ is the
number of replicas, 
\begin{equation}
b_{r}=\cosh ^{n}\left( \beta J\right) \tanh ^{r}\left( \beta J\right) ,
\end{equation}
and the trace is taken over the set of replica spin variables $\left\{
\sigma _{\alpha }\right\} $. The parameter $q_{\alpha _{1},...,\alpha _{r}}$
is the expected value of the product $\sigma _{\alpha _{1}}...\sigma
_{\alpha _{r}}$. The global order parameter $G\left( \left\{ \sigma _{\alpha
}\right\} \right) $ is defined as 
\begin{equation}
G\left( \left\{ \sigma _{\alpha }\right\} \right)
=c\sum_{r=0}^{n}\sum_{\left( \alpha _{1},...,\alpha _{r}\right)
}b_{r}q_{\alpha _{1},...,\alpha _{r}}\sigma _{\alpha _{1}}...\sigma _{\alpha
_{r}}.
\end{equation}

In the $n\rightarrow 0$ limit, the minimization of this variational free
energy with respect to the set of variables $\left\{ q_{\alpha
_{1},...,\alpha _{r}}\right\} $ leads to the stationary conditions 
\begin{equation}
q_{\alpha _{1},...,\alpha _{r}}=\frac{1}{Z_{G}}\int_{-\infty }^{+\infty
}dHp_{H}\left( H\right) \mbox{Tr}_{\sigma }\sigma _{\alpha _{1}}...\sigma
_{\alpha _{r}}\exp \left[ G\left( \left\{ \sigma _{\alpha }\right\} \right)
+\beta H\sum_{\alpha =1}^{n}\sigma _{\alpha }\right] ,
\end{equation}
where 
\begin{equation}
Z_{G}=\int_{-\infty }^{+\infty }dHp_{H}\left( H\right) \mbox{Tr}_{\sigma
}\exp \left[ G\left( \left\{ \sigma _{\alpha }\right\} \right) +\beta
H\sum_{\alpha =1}^{n}\sigma _{\alpha }\right] .
\end{equation}

In the context of the replica-symmetric Ansatz, we define an effective field 
$h$, associated with an effective probability distribution $p\left( h\right) 
$, which is equally applied on all of the replica spin variables. We then
write 
\begin{equation}
q_{\alpha _{1},...,\alpha _{r}}=\int_{-\infty }^{+\infty }dhp\left( h\right)
\tanh ^{r}\left( \beta J\right) ,
\end{equation}
from which we have 
\begin{equation}
p\left( h\right) =\int_{-\infty }^{+\infty }dHp_{H}\left( H\right)
\int_{-\infty }^{+\infty }\frac{dy}{2\pi }\exp \left[ -iy\left( h-H\right)
+G\left( \frac{iy}{\beta }\right) -c\right] .
\end{equation}

Depending on the quantity of interest, it may be more convenient to work
either with the global order parameter $G$ or with the effective probability
distribution $p$. In the replica-symmetric context, it is easy to see that 
\begin{equation}
G\left( \left\{ \sigma _{\alpha }\right\} \right) =G\left( \sum_{\alpha
=1}^{n}\sigma _{\alpha }\right) .
\end{equation}
Therefore, taking into account the double-delta distribution (\ref{doubled}%
), the extremization of the variational free energy is reduced to the
problem of searching the solutions of the non-linear integral equation 
\begin{equation}
p\left( h\right) =\int_{-\infty }^{+\infty }\frac{dy}{2\pi }\exp \left[
-iyh+\ln \cosh \left( iyH_{R}\right) -c+G\left( \frac{iy}{\beta }\right) %
\right] ,  \label{integr}
\end{equation}
where 
\begin{equation}
G\left( y\right) =c\int_{-\infty }^{+\infty }dxp\left( x\right) \exp \left\{
y\tanh ^{-1}\left[ \tanh \left( \beta J\right) \tanh \left( \beta x\right) %
\right] \right\} .
\end{equation}

In the context of the replica-symmetric Ansatz, it is known that the
Viana-Bray spin-glass problem, with a symmetric distribution of exchange
interactions and no external fields, can also be formulated in terms of a
similar integral equation for the distribution of the effective fields \cite
{Kanter1987}\cite{Mezard1987}. In this spin-glass case, if we restrict to
the analysis of the ground state ($\beta \rightarrow \infty $), it is
possible to write an analytic solution as a sum of delta functions peaked at
integer multiples of the variance $J$ of the distribution of exchanges. In
the present case, however, a similar solution requires the additional
assumption that $H_{R}/J$ is restricted to the set of integer numbers. Under
these conditions, in the ground state, the integral equation (\ref{integr})
can be exactly solved in terms of sums of delta functions.

In the ground state ($\beta \rightarrow \infty $), it is possible to show
that 
\begin{equation*}
G\left( \frac{iy}{\beta }\right) =c\int_{-\infty }^{-J}dxp\left( x\right)
\exp \left( -iyJ\right) +c\int_{-J}^{+J}dxp\left( x\right) \exp \left(
ixy\right) +
\end{equation*}
\begin{equation}
+c\int_{+J}^{\infty }dxp\left( x\right) \exp \left( iyJ\right) .
\end{equation}
Assuming that $H_{R}=\omega J$, with $\omega =0,1,2,...$, we can also write 
\begin{equation}
G\left( \frac{iy}{\beta }\right) =A+B\exp \left( iyJ\right) +C\exp \left(
-iyJ\right) ,
\end{equation}
with 
\begin{equation}
c=A+B+C,  \label{coef1}
\end{equation}
\begin{equation}
\frac{1}{c}A=\frac{1}{2}e^{\left( A-c\right) }\left[ \left( \frac{C}{B}%
\right) ^{\omega /2}+\left( \frac{C}{B}\right) ^{-\omega /2}\right]
I_{\omega }\left( 2\sqrt{BC}\right) ,  \label{coef2}
\end{equation}
and 
\begin{equation*}
\frac{1}{c}B=1-\frac{C}{2}e^{\left( A-c\right) }\int_{0}^{1}dte^{\left(
tC\right) }\{\left[ \frac{C\left( 1-t\right) }{B}\right] ^{\frac{\omega -1}{2%
}}I_{\omega -1}\left( 2\sqrt{BC\left( 1-t\right) }\right) +
\end{equation*}
\begin{equation}
+\left[ \frac{C\left( 1-t\right) }{B}\right] ^{-\frac{\omega +1}{2}%
}I_{\omega +1}\left( 2\sqrt{BC\left( 1-t\right) }\right) \},  \label{coef3}
\end{equation}
for $ \omega\geq 1 $, where $I_{\omega }\left( x\right) $ is the modified Bessel function. Both
ferromagnetic ($B\neq C$) and paramagnetic ($B=C$) solutions are represented
by this expression. The effective probability distribution $p$ is written as
a sum of delta functions, 
\begin{equation}
p\left( h\right) =\sum_{k=-\infty }^{+\infty }a_{k}\delta \left( h-kJ\right)
,
\end{equation}
where 
\begin{equation}
a_{k}=\frac{1}{2}\exp\left(A-c\right)\left[ \left( \frac{B}{C}\right) ^{\frac{k+\omega }{2}%
}I_{k+\omega }\left( 2\sqrt{BC}\right) +\left( \frac{B}{C}\right) ^{\frac{%
k-\omega }{2}}I_{k-\omega }\left( 2\sqrt{BC}\right) \right] .
\end{equation}

\section{Analysis of stability}

The analysis of stability of the replica-symmetric solutions is based on the
investigation of the eigenvalues of the Hessian matrix associated with the
variational free energy, given by Eq. (\ref{free}), which can be rewritten
as 
\begin{equation*}
\beta f\left[ G\right] -\frac{c}{2}=\sum_{r=0}^{n}\sum_{\left( \alpha
_{1},...,\alpha _{r}\right) }\frac{2^{-2n}}{2cb_{r}}\left[ \mbox{Tr}_{\sigma
}\sigma _{\alpha _{1}}...\sigma _{\alpha _{r}}G\left( \left\{ \sigma
_{\alpha }\right\} \right) \right] ^{2}-
\end{equation*}
\begin{equation}
\ln \mbox{Tr}_{\sigma }\exp \left[ G\left( \left\{ \sigma _{\alpha }\right\}
\right) +\ln \cosh \left( \beta H_{R}\sum_{\alpha =1}^{n}\sigma _{\alpha
}\right) \right] .
\end{equation}
We then write 
\begin{equation}
\mbox{Tr}_{\tau }\frac{\delta ^{2}\beta f\left[ G\right] }{\delta G\left(
\left\{ \sigma _{\alpha }\right\} \right) \delta G\left( \left\{ \tau
_{\alpha }\right\} \right) }\varphi \left( \left\{ \tau _{\alpha }\right\}
\right) =\lambda \varphi \left( \left\{ \sigma _{\alpha }\right\} \right) ,
\end{equation}
which can be cast in the form 
\begin{equation*}
\varphi \left( \left\{ s_{\alpha }\right\} \right) =c\lambda \mbox{Tr}_{\tau
}\exp \left[ \beta J\sum_{\alpha =1}^{n}\tau _{\alpha }s_{\alpha }\right]
\varphi \left( \left\{ \tau _{\alpha }\right\} \right) +
\end{equation*}
\begin{equation*}
+\frac{c}{Z_{G}}\mbox{Tr}_{\tau }\exp \left[ G\left( \widehat{\tau }\right)
+\ln \cosh \left( \beta H_{R}\widehat{\tau }\right) +\beta J\sum_{\alpha
=1}^{n}\tau _{\alpha }s_{\alpha }\right] \varphi \left( \left\{ \tau
_{\alpha }\right\} \right) -
\end{equation*}
\begin{equation}
\frac{1}{Z_{G}}G\left( \widehat{s}\right) \mbox{Tr}_{\tau }\exp \left[
G\left( \widehat{\tau }\right) +\ln \cosh \left( \beta H_{R}\widehat{\tau }%
\right) \right] \varphi \left( \left\{ \tau _{\alpha }\right\} \right) ,
\label{eigen}
\end{equation}
where 
\begin{equation}
\widehat{\tau }=\sum_{\alpha =1}^{n}\tau _{\alpha },\qquad \widehat{s}%
=\sum_{\alpha =1}^{n}s_{\alpha }.
\end{equation}
In the $n\rightarrow 0$ limit, it is easy to show that there is a constant
eigenvector, $\varphi \left( \left\{ s_{\alpha }\right\} \right) =%
\mbox{constant}$, with $1/c$ as the associated eigenvalue.

According to the work of De Dominicis and Mottishaw \cite{Dedominicis1987},
in the context of the replica-symmetric approximation the space of $2^{n}$
eigenvectors is spanned by a set of eigenvectors parameterized by functions
of two variables, of the form 
\begin{equation}
\varphi \left( \left\{ \sigma _{\alpha }\right\} \right) =\varphi _{\left\{
\mu _{\alpha }\right\} }\left( \widehat{\sigma };q_{\sigma \mu }\right) ,
\end{equation}
where 
\begin{equation}
\widehat{\sigma }=\sum_{\alpha =1}^{n}\sigma _{\alpha },\qquad q_{\sigma \mu
}=\sum_{\alpha =1}^{n}\sigma _{\alpha }\mu _{\alpha },
\end{equation}
and the spin configuration $\left\{ \mu _{\alpha }\right\} $ is used to
label the eigenvectors. From Eq. (\ref{eigen}), in the $n\rightarrow 0$
limit, we derive an integral equation for the eigenvalues, 
\begin{equation*}
\varphi _{\mu }\left( x,y\right) =-\frac{G\left( x\right) }{\exp \left(
c\right) }\iint_{-\infty }^{+\infty }dmdr\iint_{-\infty }^{+\infty }\frac{%
dudv}{\left( 2\pi \right) ^{2}}\left[ \frac{\cosh \left( \beta u+\beta
v\right) }{\cosh \left( \beta u-\beta v\right) }\right] ^{\mu /2}\times
\end{equation*}
\begin{equation*}
\exp \left[ G\left( \frac{im}{\beta }\right) -imu-irv+\ln \cosh \left(
iH_{R}m\right) \right] \varphi _{\mu }\left( \frac{im}{\beta },\frac{ir}{%
\beta }\right) +
\end{equation*}
\begin{equation*}
c\iint_{-\infty }^{+\infty }dmdr\iint_{-\infty }^{+\infty }\frac{dudv}{%
\left( 2\pi \right) ^{2}}\left\{ \lambda +\exp \left[ G\left( \frac{im}{%
\beta }\right) +\ln \cosh \left( iH_{R}m\right) -c\right] \right\}
\end{equation*}
\begin{equation*}
\times \exp \left( imu+irv\right) \left[ \frac{\cosh \left( \beta J+\beta
u+\beta v\right) }{\cosh \left( \beta J-\beta u-\beta v\right) }\right] ^{%
\frac{x+y}{4}}
\end{equation*}
\begin{equation*}
\times \left[ \cosh \left( \beta J+\beta u+\beta v\right) \cosh \left( \beta
J-\beta u-\beta v\right) \right] ^{\frac{\mu }{4}}\left[ \frac{\cosh \left(
\beta J+\beta u-\beta v\right) }{\cosh \left( \beta J-\beta u+\beta v\right) 
}\right] ^{\frac{x-y}{4}}
\end{equation*}
\begin{equation}
\times \left[ \cosh \left( \beta J+\beta u-\beta v\right) \cosh \left( \beta
J-\beta u+\beta v\right) \right] ^{-\frac{\mu }{4}}\varphi _{\mu }\left( 
\frac{im}{\beta },\frac{ir}{\beta }\right) .
\end{equation}

First, we find the longitudinal eigenvalues, in other words, the eigenvalues
in the subspace spanned by $\varphi _{\left\{ \mu _{\alpha }\right\} }\left( 
\widehat{\sigma };q_{\sigma \mu }\right) =\varphi \left( \widehat{\sigma }%
\right) $. In the ground state, it is not difficult to see that these
longitudinal eigenvectors can be written as 
\begin{equation}
\varphi _{L}\left( \frac{ix}{\beta }\right) =A_{L}+B_{L}\exp \left(
ixJ\right) +C_{L}\exp \left( -ixJ\right) .
\end{equation}
The problem reduces to the calculation of the eigenvalues of a $3\times 3$
matrix, which are given by $\lambda _{1}=1/c$, associated with the constant
eigenvector, and 
\begin{equation}
\lambda _{2,3}=\frac{1}{c}\left( 1-A\right) \pm D,  \label{ll1}
\end{equation}
where $A=ca_{0}$ is given by Eq. (\ref{coef2}), and $D=\left(
a_{1}a_{-1}\right) ^{1/2}$ is given by the expression 
\begin{equation*}
D=\frac{1}{2}\exp \left( A-c\right) \{\left( \frac{C}{B}\right) ^{\omega
}I_{\omega +1}\left( 2\sqrt{BC}\right) I_{\omega -1}\left( 2\sqrt{BC}\right)
+I_{\omega -1}^{2}\left( 2\sqrt{BC}\right) +
\end{equation*}
\begin{equation}
+I_{\omega +1}^{2}\left( 2\sqrt{BC}\right) +\left( \frac{C}{B}\right)
^{-\omega }I_{\omega +1}\left( 2\sqrt{BC}\right) I_{\omega -1}\left( 2\sqrt{%
BC}\right) \}^{1/2},
\end{equation}
with the coefficients $A$, $B$, and $C$, given by Eqs. (\ref{coef1})-(\ref
{coef3}). These longitudinal eigenvalues, however, lead to familiar
mean-field results. At small values of the random field, the paramagnetic
solution is unstable, while a ferromagnetic solution is stable. At large
values of the random field, there is only a (stable) paramagnetic solution
as in the case of spin glasses on the Bethe lattice (note that, at
sufficiently large random fields, that is, for $H_{R}\rightarrow \infty $,
we have $A,D\rightarrow 0$). The critical border separating these
paramagnetic and ferromagnetic phase is of order $cJ$ (see Table 1 for
specific values\ ).

We now turn to the eigenvalues associated with the transversal sector. For $%
\mu =0$, and an eigenvector of the form 
\begin{equation*}
\varphi _{\mu =0}\left( \frac{ix}{\beta },\frac{iy}{\beta }\right)
=A_{0}+B_{1+}\exp \left( ixJ\right) +B_{1-}\exp \left( -ixJ\right) +
\end{equation*}
\begin{equation*}
+B_{2+}\exp \left( iyJ\right) +B_{2-}\exp \left( -iyJ\right) +C_{++}\exp
\left( ixJ+iyJ\right) +
\end{equation*}
\begin{equation}
+C_{+-}\exp \left( ixJ-iyJ\right) +C_{-+}\exp \left( -ixJ+iyJ\right)
+C_{--}\exp \left( -ixJ-iyJ\right) ,  \label{muzero}
\end{equation}
the problem is reduced to the calculation of the eigenvalues of a $9\times 9$
matrix. The nine eigenvalues of this transversal sector are given by $%
\lambda _{T1}=1/c$, associated with a constant eigenvector, 
\begin{equation}
\lambda _{T2,T3}=\frac{1}{c}\left( 1-A\right) ,  \label{lt1}
\end{equation}
\begin{equation}
\lambda _{T4,T5,T6}=\frac{1}{c}\left( 1-A\right) +D,  \label{lt2}
\end{equation}
and 
\begin{equation}
\lambda _{T7,T8,T9}=\frac{1}{c}\left( 1-A\right) -D,  \label{lt3}
\end{equation}
which should be compared with Eq. (\ref{ll1}) for the non-trivial
eigenvalues of the longitudinal sector. In contrast to the spin-glass case,
the eigenvalues of this transverse sector do not lead to any additional
instability. According to a numerical analysis of these eigenvalues, the
replica-symmetric paramagnetic solution remains stable for sufficiently
large random fields and the ferromagnetic solution is stable in its region
of existence (see Table 1).

The analysis of the transversal sector with $\mu \neq 0$ can be carried out
with same Ansatz, 
\begin{equation}
\varphi _{\mu \neq 0}\left( \frac{ix}{\beta },\frac{iy}{\beta }\right)
=\sum_{\theta \in \mathbf{Z}}\exp \left( \frac{1}{2}im\theta J\right)
\varphi _{\theta,\mu =0}\left( \frac{ix}{\beta },\frac{iy}{\beta }\right) .
\end{equation}
Although the secular matrix becomes infinite, it is easy to see that the
eigenvalues are still given by the same expressions of Eqs. (\ref{lt1})-(\ref
{lt3}). Again, we conclude that the replica-symmetric solution is stable in
the presence of sufficiently large random fields.

In Table 1, we list the numerical solutions for the smallest eigenvalue,
given by equation (\ref{lt3}), with $c=3,4,5,$ and $6$, for the paramagnetic
(pm) and ferromagnetic (fm) solutions. There is no simultaneous instability
of both solutions and thus no indication of breaking of replica symmetry. Note
that the dashes in this table correspond to the absence of a ferromagnetic
solution (in which case the paramagnetic solution is stable).\bigskip

\begin{center}
Table 1\bigskip

\begin{tabular}{lllllll}
$H_{R}/J$    & $0$       & $1$       & $2$       & $3$      & $4$      & $c$ \\
             &           &           &           &          &          &     \\ 
$\lambda pm$ & $-0.1798$ & $-0.0661$ & $0.1030$  & $0.2253$ & $0.2919$ & $3$ \\ 
$\lambda fm$ & $0.2738$  & $0.1157$  & $---$     & $---$    & $---$    & $3$ \\ 
             &           &           &           &          &          &     \\
$\lambda pm$ & $-0.1876$ & $-0.1159$ & $0.0129$  & $0.1212$ & $0.1906$ & $4$ \\ 
$\lambda fm$ & $0.2302$  & $0.1669$  & $---$     & $---$    & $---$    & $4$ \\ 
             &           &           &           &          &          &     \\
$\lambda pm$ & $-0.1876$ & $-0.1377$ & $-0.0369$ & $0.0578$ & $0.1258$ & $5$ \\ 
$\lambda fm$ & $0.1930$  & $0.1666$  & $0.0819$  & $---$    & $---$    & $5$ \\ 
             &           &           &           &          &          &     \\
$\lambda pm$ & $-0.1847$ & $-0.1476$ & $-0.0667$ & $0.0161$ & $0.0808$ & $6$ \\ 
$\lambda fm$ & $0.1642$  & $0.1529$  & $0.1153$  & $---$    & $---$    & $6$%
\end{tabular}
\end{center}

\section{Conclusions}

We have investigated the stability of the replica-symmetric solutions of a
random-field Ising ferromagnet on a lattice of finite mean connectivity. At
low temperatures and for sufficiently large random fields, the analysis of
the eigenvalues of the Hessian matrix associated with the variational free
energy leads to stable replica-symmetric paramagnetic solutions (at smaller
random fields, the replica-symmetric ferromagnetic solution is stable). The
present calculations do not support the existence of a glassy phase, as
suggested by earlier proposals \cite{Dealmeida1987}\cite{Pastor2002}\cite
{Mezard1992}. However, a more detailed analysis of the phase diagram, in
terms of field and temperature, still demands considerable work, including
both analytical and refined numerical calculations. The assumption of a
discrete distribution of effective fields, which was written as a sum of
delta functions, may not capture the subtleties of the glassy behavior. As
in the spin-glass case, we cannot rule out the existence of field-induced
glassy and mixed ferromagnetic-glassy phases.

It is interesting to point out a connection with the recent work by Pastor
and collaborators \cite{Pastor2002}. A truncation of the variational free
energy, given by Eq. (\ref{free}), leads to a ``high-temperature
approximation,''\ 
\begin{equation*}
f_{app}=\frac{1}{\beta }\lim_{n\rightarrow 0}\frac{1}{n}\{\frac{c}{2}\tanh
\left( \beta J\right) \sum_{\alpha }m_{\alpha }^{2}+\frac{c}{2}\left[ \tanh
\left( \beta J\right) \right] ^{2}\sum_{\alpha <\beta }q_{\alpha \beta }^{2}-
\end{equation*}
\begin{equation}
-\ln \left[ \int_{-\infty }^{+\infty }dHp_{H}\left( H\right) Z_{app}\right]
\},
\end{equation}
where 
\begin{equation*}
Z_{app}=\mbox{Tr}_{\sigma }\exp \{c\tanh \left( \beta J\right) \sum_{\alpha
}\sigma _{\alpha }+
\end{equation*}
\begin{equation}
+c\left[ \tanh \left( \beta J\right) \right] ^{2}\sum_{\alpha <\beta
}q_{\alpha \beta }\sigma _{\alpha }\sigma _{\beta }+\beta H\sum_{\alpha
=1}^{n}\sigma _{\alpha }\}.
\end{equation}
The results of Pastor et al.\cite{Pastor2002} for the paramagnetic phase are
recovered if we introduce the additional approximation $\tanh \beta J=\beta
J+O\left[ \left( \beta J\right) ^{3}\right] $, and discard higher-order
terms. In this approximation, for a Gaussian distribution of random fields
with variance $H_{R}$, the replica-symmetric paramagnetic phase is unstable
along the field axis.

\textbf{Acknowledgments}

We thank the financial support of the Brazilian agencies FAPESP, CNPq, and
CAPES.

\end{document}